\documentclass[aps,twocolumn,noshowpacs,superscriptaddress,floatfix,nofootinbib]{revtex4-1}

\usepackage[utf8]{inputenc}
\usepackage{quantikz}
\usepackage{amsmath,amssymb}
\usepackage{amsthm,algorithmic,epsfig,tikz}
\usepackage{algorithm}
\usepackage{graphicx}
\usepackage{caption}
\usepackage{subcaption}
\usepackage{hyperref}  
\hypersetup{colorlinks,citecolor=black,urlcolor=black,linkcolor =black,hypertexnames=true}
\newcommand{\sumthree}{\operatorname*{\sum\sum\sum}}
\begin{document}

\title{Conditional Generative Models for Learning Stochastic Processes}
\author{Salvatore Certo}
\affiliation{Deloitte Consulting, LLP}

\author{Anh Pham}
\affiliation{Deloitte Consulting, LLP}

\author{Nicolas Robles}
\affiliation{IBM Quantum, IBM Research}

\author{Andrew Vlasic}
\affiliation{Deloitte Consulting, LLP}

\date{\today}

\begin{abstract} 
A framework to learn a multi-modal distribution is proposed, denoted as the Conditional Quantum Generative Adversarial Network (C-qGAN). The neural network structure is strictly within a quantum circuit and, as a consequence, is shown to represent a more efficient state preparation procedure than current methods.  This methodology has the potential to speed-up algorithms, such as Monte Carlo analysis. In particular, after demonstrating the effectiveness of the network in the learning task, the technique is applied to price Asian option derivatives, providing the foundation for further research on other path-dependent options. 

\end{abstract}

{
\let\clearpage\relax
\maketitle
}

\section{Introduction}
The work in this manuscript establishes a framework for conditional quantum generative models to learn conditional distributions, such as in stochastic processes conditioned on time. Quantum machine learning (QML) is leveraged to train the generative model with a specific variational circuit that offers a compact structure, yielding an efficient state preparation procedure that can be utilized as a subprocess in various algorithms. An example of pricing the derivatives of Asian options is given. 

Many algorithms have been created to provide a quadratic speedup of certain calculations \cite{montanaro}, including those in finance,such as Monte Carlo simulations for derivative pricing and Value at Risk (VaR).  These algorithms involve applying functions to one or more random variables, and extracting the expected value of the function which is the quantity of interest.  For example, finding the fair price for a European Call Option includes evaluating the payoff function based on the strike price and the distribution of the expected spot prices for the asset at maturity.  In a quantum setting, this example would involve several steps, including loading the spot price distribution into a quantum state, computing the payoff function, and finally extracting the expected value from the amplitude of a quantum state \cite{egger2019credit, W19, S19, O19}.  

While demonstrations on pricing these simple derivatives, which are contracts contingent on the derived performance of the underlying asset, using quantum algorithms has shown promise but  it is unclear what the best methodology and implementation would be for extending these same methodologies to more complex financial instruments, while retaining the quadratic advantage. It is known that state preparation and quantum arithmetic are two of the more expensive operations on quantum devices. For example, the well-known Grover-Rudolph method \cite{grover2002creating} has been shown to negate a quadratic speed-up \cite{herbert2021problem}. These negations are exacerbated when attempting to approximate stochastic processes, where the evolution of underlying dynamics across many time steps effects the expectation values of interest. Further literature of dealing with stochastic differential equations (SDEs) and the partial differential equation arising from these SDEs can be found in \cite{quboendo, fk}.

Notably, prior work on pricing instruments whose value is derived from a stochastic process has suggested separate registers for each time step \cite{S19} and expensive quantum operations to evolve the process arithmetically.  This work takes a different approach, which is aimed towards learning the dynamics of the stochastic process with QML in tandem with a control register that allows for efficient operations on specific timesteps of that process.  With this intuition, we explore the use of general Conditional Quantum Generative Models (C-qGM), and a specific implementation of them in our hybrid Conditional Quantum Generative Adversarial Networks (C-qGAN) algorithm. The C-qGAN was inspired from similar frameworks in the classical setting \cite{mirza2014conditional}. Similar approach of using a C-qGAN was also explored by Liu et al \cite{liu2021hybrid} where the generative algorithm was applied to generate simple stripes images with uniform distribution. Within this manuscript, we applied the C-qGAN algorithm to generate multiple non-uniform distributions as a mechanism to load these distributions within a process of evaluating Asian option. Furthermore,  the implementation of pricing an Asian option is explicitly provided, and ran on a quantum simulator with 13 qubits. The results of this implementation are discussed.

\section{Generative Models}

Generative models play an important role in many applications, for instance chemistry, text generation, and finance.  Many of these models are trained adversarially using the Generative Adversarial Network model (GAN), where two deep neural networks compete against each other to either decipher whether data is synthetic or to create synthetic data indecipherable from real-data \cite{creswell2018generative}. In particular, the discriminator is noted as the network that learns to discern real data from fake data, and the other network, noted as the generator, learns how to mimic real data in order to fooling the discriminator.

As one may observe, GANs are useful when the distribution of the random variable is unknown or is of increasingly high dimension to model. However, if the distribution of the random variable is known, or of small dimension, the discriminator in the adversarial component can be dropped and the generator can trained explicitly with the known distribution at hand. 

Recently, the network structure of a GAN has been extended to quantum circuits, and intuitively called a quantum generative adversarial network (qGAN) \cite{lloyd2018quantum, Zoufal_2019,chakrabarti2019quantum, situ2020quantum}. Initial numerical results have shown that QML models can be trained with less data than their classical counterparts \cite{caro2021generalization}, displaying a quantum advantage. 

\subsection{Quantum Generative Adversarial Networks}

The goal of any state preparation procedure is to load a classical data-point into a quantum state
$$
\ket{\psi}_{n} = \sum_{i=0}^{2^{n}-1} \sqrt{p_{i}}\ket{i}_{n},
$$
where $\ket{i}$ is the binary string of the integer $i$, which represents the event of being to the left or right of the $i$ number with probability $p_{i}$. Observe that, in practice, this quantum state is usually an approximated discretized distribution of a continuous distribution.

It has been shown that directly loading arbitrary states into a quantum system requires a number of gates with time complexity of $\mathcal{O}(2^n)$  \cite{plesch2011quantum, sanders2019black, herbert2021problem}. However, there have been advancements, such as Gonzalez-Conde et al.\cite{gonzalez2023quantum} who made an improvement to the Grover-Rudolph algorithm, decreasing the number of gates to $\mathcal{O}(2^{k_0(\epsilon)})$, where $\epsilon$ is the infidelity respect to the exact state and the parameter $k_0(\epsilon)$ asymptotically independent of $n$. 

There is potential to derive a quantum subprocess to load in arbitrary states that reduces the number of gates to size $\mathcal{O}(\mbox{poly}(n))$ by reducing many of the controlled gates in a general circuit \cite{Zoufal_2019}. 

\begin{figure}[h!]
    \centering
    \includegraphics[width=250px]{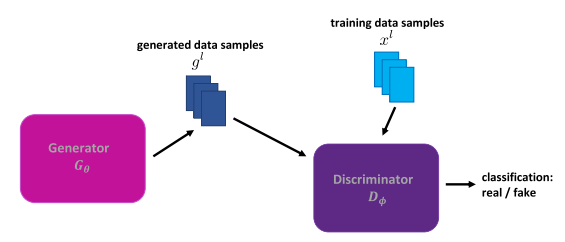}
    \caption{General structure of a qGAN as given in Zoufal et al. \cite{Zoufal_2019}.}
    \label{fig:qGAN}
\end{figure}

This subprocess is typically a qGAN. In general, the deep neural network structure of a classical generator is replaced with a parameterized quantum circuit in such a manner as to reduce the amount of data records to train a robust statistical model that mimics the unknown distribution more accurately and with less training time; see Figure \ref{fig:qGAN} for a high-level structure for training. These distributions, as noted above, create the desired quantum state. A high-level depiction of the general structure of a qGAN is shown in Figure \ref{fig:qGAN}.

Using the reduced general framework of a qGAN, it is proposed to utilize the simple QML neural network architecture as a basis to derive a novel temporal based distribution, thereby retaining the quantum-advantage in applications which require these distributions as a sub-circuit. While promising, it is not only necessary for this QML approach to limit the size of the parameterized circuit (in two-qubit gates, controlled gates, and depth), but also in reducing the training time.

\subsection{Conditional-qGANs}

\begin{figure}[h]
    \centering
    \includegraphics[width=240px]{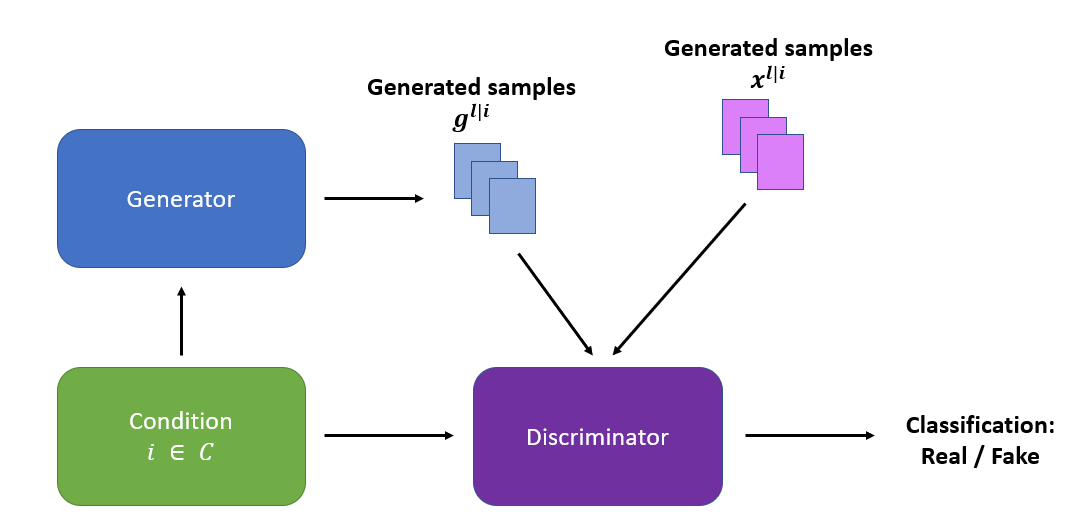}
    \caption{Proposed structure of the Conditional-qGAN.}
    \label{fig:C-qGAN}
\end{figure}

Although qGANs are useful for modeling intricate  high-dimensional distributions, qGANs assume the underlying distribution is invariant. Classically, a multi-modal distribution has been algorithmically modeled as a conditionally generative adversarial network \cite{mirza2014conditional,odena2016semi}. For the auxiliary information, $\mathbf{y}$, the objective function has the form 
\begin{equation*}
    \begin{split}
        \min_{G}\max_{D}V(G,D) & = \mathbb{E}_{x \sim p_{data}(\mathbf{x}) }\Big[ \log\big(D(\mathbf{x}|\mathbf{y} \big) \Big] \\
        & + \mathbb{E}_{z \sim p_z(\mathbf{z})}\Big[ \log\left(1-D\big( G(\mathbf{z}|\mathbf{y}) \right) \big) \Big].
    \end{split}
\end{equation*}

Following the method in \cite{mirza2014conditional} to extract information from a multi-modal distribution, we derive both a classical-quantum hybrid qGAN, denoted as the Conditional-qGAN (C-qGAN) and a conditional quantum generative model (C-qGEN). Figure \ref{fig:C-qGAN} shows the high-level structure for training. The hybrid algorithm follows the work of Liu et al. \cite{liu2021hybrid}, where only the generator is strictly in a quantum circuit and conditional information is fed by a separate qubit register. Zoufal et al. \cite{Zoufal_2019} with a traditional GAN displayed robustness with this hybrid architecture. Jayasinha \cite{PJ2021} created a hybrid conditional GAN package. 

\begin{figure*}[ht!]
  \centering
    \begin{subfigure}[b]{0.9\textwidth}
    \centering
    \includegraphics[width=450pt]{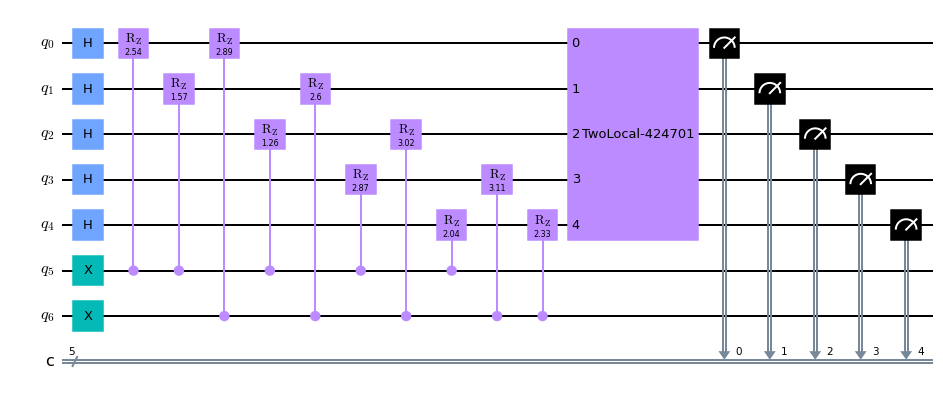} 
    \caption{The circuit consists of a register circuit from qubits $q_0$ through $q_4$, where the outputs form the discretized distribution with $2^{5}$ states, and an ancilla register, where qubits $q_5$ and $q_6$ which are the conditional qubits.}
    \label{fig:c_distloader_circ}
    \end{subfigure}
    \begin{subfigure}[b]{0.8\textwidth}
    \centering
    \includegraphics[width=320pt]{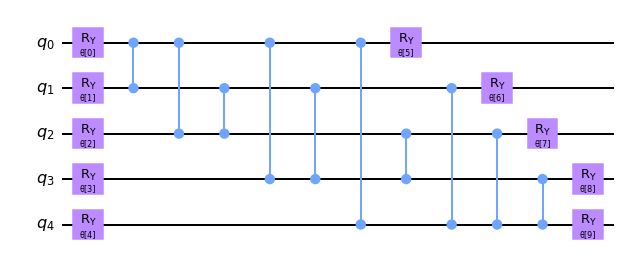}
    \caption{This figure is the decomposition of the Two-Local subprocess, created with two layers of $Ry$ gates and a layer of one-to-all control-$Z$ gates in-between the $Ry$ layers.}
    \label{fig:c_distloader_twoloc}
    \end{subfigure}
\caption{Example of a Conditional Generative Model with learned parameters for four initialized conditions.}
\label{fig:c_distloader}
\end{figure*}

To make the hybrid algorithm more tangible, we follow the SGAN algorithm in \cite{odena2016semi} and Liu et al. \cite{liu2021hybrid}, with the process described in Algorithm \ref{alg:gan}. The SGAN algorithm assume a general number of classes and fake data, $[ \mbox{CLASS}_1, \mbox{CLASS}_2, \ldots, \mbox{CLASS}_n, \mbox{FAKE}]$, and hence $D$ could have $n+1$ output units. The authors call this network D/C. While the general outline in Algorithm \ref{alg:gan} follows the outline in \cite{liu2021hybrid}, the tensor architecture of the generator is significantly smaller than the generator architecture given in \cite{liu2021hybrid}. Although the example given in the paper describes a generator with a similar gate complexity; please see Figure \ref{fig:c_distloader} for an illustrative example.

\begin{algorithm}[H]
\caption{qSGAN}\label{alg:gan}
\raggedright\textbf{Input:} Conditional ansatz CMAP with $c$ $\alpha$ parameters, $I$ the total number of iterations, and $m$ the size of the minibatch
\begin{algorithmic}[1]
\STATE $i \gets 1$
\WHILE{$i \leq I$}
    \STATE \hspace{\algorithmicindent} Draw $m$ noise samples $\{z^{(1)}, \ldots, z^{(m)} \}$ from prior $p_{G}(z)$
    \STATE \hspace{\algorithmicindent} Draw $m$ examples $\{ (x^{(1)},y^{(1)}),(x^{(2)},y^{(2)}), \ldots, (x^{(m)},y^{(m)}) \}$ from data $p_{d}(x)$
    \STATE \hspace{\algorithmicindent} Apply gradient descent on the parameters of $D$ with the minibatch size $2m$
    \STATE \hspace{\algorithmicindent} Draw $m$ noise samples $\{z^{(1)}, \ldots, z^{(m)} \}$ from prior $p_{G}(z)$
    \STATE \hspace{\algorithmicindent} Apply gradient descent on the parameters of $G$ with the minibatch size $m$
    \STATE \hspace{\algorithmicindent} $i \gets i+1$
\ENDWHILE
\end{algorithmic}
\end{algorithm}

For the C-qGEN we derive a generative algorithm that directly transfers classical information into a quantum circuit. Although the transfer of information is conducted within a variational subprocess, the loss-function only requires the distribution. Thus, simplifying the variational algorithm.   

\begin{algorithm}[H]
\caption{Conditional Generative Model Training}\label{alg:bsw}
\raggedright\textbf{Input:} Conditional ansatz CMAP with $c$ $\alpha$ parameters, distribution ansatz $Ansatz_{D}$ with $n$ $\theta$ parameters.  $T$ conditions and specific conditions $t$ mapped to binary integers to be loaded on condition register of size $c \leq 2^{T}$. Desired precision $p$
\begin{algorithmic}[1]
\STATE $Loss \gets 0$, $\epsilon \gets T - Loss$
\STATE \textbf{while} $\epsilon >= p$
    \STATE \hspace{\algorithmicindent} \textbf{For} $t = 0$ to $T$ \textbf{Do}
        \STATE \hspace{\algorithmicindent} Initialize $\ket{t}$
        \STATE \hspace{\algorithmicindent} Measure $Ansatz_{D}(\theta_{n})$
        \STATE \hspace{\algorithmicindent} Loss += Loss-Function
        \STATE \hspace{\algorithmicindent} $\epsilon \gets T - \mbox{Loss}$
    \STATE Classical-Optimizer$(\alpha_{c}, \theta_{n}, \mbox{Loss})$
\STATE \textbf{return} $\epsilon$
\end{algorithmic}
\end{algorithm}

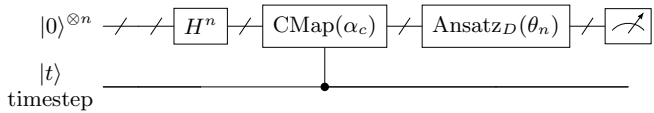
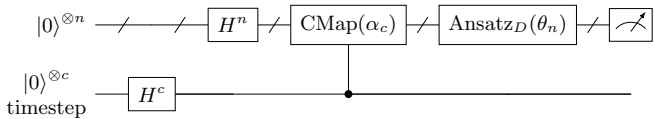
\begin{figure}	
	\noindent
	\begin{subfigure}[b]{0.5\textwidth}
		\noindent \resizebox{\columnwidth}{!}{
    \begin{quantikz}[thin lines] 
        \lstick{$\ket{0}^{\otimes n}$} & \qwbundle[alternate]{} & \gate{H^n} \qwbundle[alternate]{} & \gate[wires=1]{\operatorname{CMap} (\alpha_c)} \qwbundle[alternate]{} & \gate{\textnormal{Ansatz}_D (\theta_n)} \qwbundle[alternate]{} &\meter{}\qwbundle[alternate]{}
        \\ \lstick{$\ket{t}$\\timestep} & \qw  & \qw & \ctrl{-1}  & \qw & \qw
    \end{quantikz}   }
		\caption{General circuit structure for loading $T$ timesteps on one register with variational parameters $\alpha$ and $\theta$.}
\label{fig:cir} 	
	\end{subfigure}
	\begin{subfigure}[b]{0.5\textwidth}
		\noindent \resizebox{\columnwidth}{!}{
    \begin{quantikz}[thin lines] 
        \lstick{$\ket{0}^{\otimes n}$} & \qwbundle[alternate]{} & \gate{H^n} \qwbundle[alternate]{} & \gate[wires=1]{\operatorname{CMap} (\alpha_c)} \qwbundle[alternate]{} & \gate{\textnormal{Ansatz}_D (\theta_n)} \qwbundle[alternate]{} &\meter{}\qwbundle[alternate]{}
        \\ \lstick{$\ket{0}^{\otimes c}$\\timestep} & \gate{H^c}  & \qw & \ctrl{-1}  & \qw & \qw
    \end{quantikz}   }
    \caption{Circuit structure for averaging joint probability distributions.} \label{fig:cir_cond}
	\end{subfigure}
	\caption{Examples of different multi-modal distributions loaded into a circuit.}\label{fig:cir_general}
\end{figure}

 In particular, a circuit implementation allows for direct control over the generated data from a single quantum register, including the application of a mixtures of conditions, such as weighted or unweighted averages. Moreover, the C-qGAN circuit may integrate a single register with a parameterized circuit, which can now output completely different distributions based on how the condition register is initialized, compared to separate registers being required for GANs or QGANs. Figure \ref{fig:c_distloader} yields an explicit implementation of this circuit where the conditional states $00$, $01$, $10$, and $11$ represent a condition. Figure \ref{fig:cir_general} shows a general circuit structure.

To display the observation of simpler gate complexity, consider the circuit in Figure \ref{fig:c_distloader}. Within this circuit, a one-dimensional distribution is approximated with 5 qubits and four sequentially conditioned distributions, and is implemented with a total of 7 qubits within the register. For this circuit, given the simplicity of the control gates and accounting for the difference of native gates with each quantum processing unit, each gate is counted as 1. Hence, with this logic, implementing $2^m$ conditions with $n$ qubits, the total number of gates is $\displaystyle (n+m \cdot n) + 2\cdot n + \sum_{i=1}^{n}(n-i) = \mathcal{O}( n^2 )$, assuming that $m < n$ since more states requires higher precision and more qubits. Therefore, this circuit has polynomial growth. 

In comparison, a direct implementation with the Grover-Rudolph technique in \cite{gonzalez2023quantum} would require one register for each distribution, in addition to the gate complexity required to implement each register, as well as the complexity of the method to feed the output of one randomly selected register into the larger algorithm.

While Algorithm \ref{alg:bsw} gives the pseudo-algorithm to train a C-qGM, the experiments in Section \ref{sec:asian} that trained the circuit in Figure \ref{fig:c_distloader} utilized cosine similarity as the loss-function and SPSA as the classical-optimizer (see \cite{spall2005introduction}). In this algorithm, each of the four conditional states were sequentially tested for loss, and the aggregated losses were used to train the parameters of the circuit. With various numerical experiments having different combinations of loss and optimization functions, cosine similarity and SPSA were shown to perform well in learning the parameters of the circuit. Section \ref{sec:results} has further details on the numerical experiments.

\section{Asian Options}\label{sec:asian}

Let $S_{t} \in T$ be a stochastic process with $T$ timesteps, and $f(S_{t})$ be a function computed on the distribution of the random variable at time $t$.  The prerequisite for further quantum computation is to load the discretized distribution of the random variable at time $t$ in one or more quantum registers, $\ket{S}_{t}$.  The most compact and efficient state preparation procedure is to utilize only one register for the distribution and another for the time step, having the form $\ket{S_{t}}\ket{t}$.
The underlying distribution of our equity stock $S=S_t=S(t)$ follows the accepted geometric Brownian motion SDE
\begin{align}
dS_t = \mu S_t dt + \sigma S_t dW_t,
\end{align}
where $\mu$ and $\sigma$ are constants known as the drift and the volatility, respectively. Furthermore $W_t=W(t)$ denotes standard Brownian motion. 

Employing Ito's lemma and setting $F(S) = \log S$, one can see that
\begin{equation}
\begin{split}
dF & = \frac{\partial F}{\partial S} dS + \frac{1}{2} \frac{\partial^2 F}{\partial S^2} (dS)^2 \\
& = \frac{1}{S} (\mu S dt + \sigma S dW_t) - \frac{1}{2} \frac{1}{S^2} \sigma^2 S^2 dt, 
\end{split}
\end{equation}
which implies that
\begin{align}
d(\log S) = \bigg(\mu - \frac{1}{2} \sigma^2 \bigg) dt + \sigma dW_t;
\end{align}
see \cite{shreve2004stochastic} for further information. Therefore, $S_t$ has closed analytic form
\begin{align}
S(t) = S_0 \exp\bigg[\bigg(\mu - \frac{1}{2} \sigma^2 \bigg) t + \sigma W_t \bigg].
\end{align}
Since $W_t$ is a one-dimensional Brownian motion, at time $t>0$,  $W_t \sim \phi \sqrt{t}$ for $\phi \sim N(0,1)$ (see \cite{platen2010numerical}). The initial condition $S_0 > 0$ is the price of the stock at $t=0$.

The specific payoff we are interested in is the \textit{Asian option}, which averages the path taken by $S_t$ during its life cycle. Ergo, denoting $K$ as the strike price and $X_T$ as the current value, which is the average of the observations, $X_T$ has the analytic form
\begin{align}
X_T  = \frac{1}{T}\int_0^T S(t) dt.
\end{align}
Therefore, the payoff for the Asian option is calculated as $\max(X_T - K,0)$; see \cite{shreve2004stochastic} for further information. 

For further illumination of pricing derivatives, consider the \textit{lookback option} with a fixed strike (see \cite{shreve2004stochastic}). The current value is the observed maximal value, 
$$
X_T  = \max_{0 \leq t\leq T} S_t,
$$
and hence the current value of the derivative is calculated as $\max(X_T-K,0)$.

For simplicity purposes, and without loss of generality, we take $T=1$ and scale other time factors accordingly. Therefore, the analytic form of the Asian options simplify to
\begin{equation}
\begin{split}
X & = X_1 = \int_0^1 S(t) dt \\
& = S_0 \int_0^1 \exp\bigg[\bigg(\mu - \frac{1}{2} \sigma^2 \bigg) t + \sigma W_t \bigg] dt.
\end{split}
\end{equation}
To simplify matters further, let us take $\mu$ and $\sigma$ such that $\mu - \tfrac{1}{2} \sigma^2 = 0$, e.g. $\mu = \tfrac{1}{32} = 0.03125 $ and $\sigma = \tfrac{1}{4} = 0.25$. We also skip $S_0$ and add it back at the end. Therefore we end up with
\begin{align}
X_1 = \int_0^1 \exp(\sigma W_t) dt \quad \textnormal{with} \quad \sigma = \frac{1}{4}.
\end{align}
Our first step is to perform a MacLaurin expansion so that
\begin{align}
X_1 = \int_0^1 \sum_{n=0}^\infty \frac{\sigma^n W^n(t)}{n!} dt.
\end{align}
Since $X_1$ converges absolutely, one is able to interchange the summation and integration, which yields the approximation
\begin{equation}
\begin{split}
X_1 &= \sum_{n=0}^\infty \frac{\sigma^n}{n!} \int_0^1 W^n(t) dt \\
&= \int_0^1 W^0(t) dt + \sigma \int_0^1 W(t) dt + \frac{\sigma^2}{2} \int_0^1 W^2(t) dt \nonumber \\
&\quad + O(W^3(t)).
\end{split}
\end{equation}

While one needs to show that the following approximation is applied with minimal information loss, we give a manner to simplify the noise, which in turn would assist in faster speed-ups of Monte Carlo analysis.  

The first term holds the following equivalency in distribution,
$$
\int_0^1 W^0(t) dt \stackrel{d}{=} \int_0^1 dt = 1. 
$$

For the linear term we employ integration by parts on a slightly more general term
\begin{equation}
\begin{split}
\int_0^T W_t dt & = tW_t \bigg|_0^T - \int_0^T tdW_t \\
& = T \int_0^T dW_t - \int_0^T t dW_t = \int_0^T (T-t)dW_t .
\end{split}
\end{equation}
Next, we use the well-known result that if $f(t)$ is a deterministic square integrable function, then the stochastic integral
\begin{align}
\int_0^T f(t) dW_t \sim N \bigg(0, \int_0^T |f(t)|^2 dt\bigg).
\end{align}
Therefore,
\begin{align}
\int_0^T W_t dt \sim N \bigg(0, \int_0^T |T-t|^2 dt\bigg) \sim N\bigg(0, \frac{T^3}{3} \bigg).
\end{align}
Recalling that $T=1$, we obtain
\begin{align}
\int_0^1 W(t) dt \sim N \bigg(0, \frac{1}{3} \bigg).
\end{align}
The quadratic term is much more interesting, as we need to make use of the Karhunen-Loeve expansion of the Wiener process
\begin{align}
W(s) = \frac{\sqrt{2}}{\pi} \sum_{k=1}^\infty Z_k \frac{\sin\Big( (k-\tfrac{1}{2})\pi s \Big)}{k-\tfrac{1}{2}},
\end{align}
where $Z_k$ are independent and identically distributed (i.i.d.) standard normal random variables. Using the orthogonality relation
\begin{align}
\int_0^1 \sin\Big( (j-\tfrac{1}{2})\pi s \Big) \sin\Big( (k-\tfrac{1}{2})\pi s \Big) ds
=
\begin{cases}
\tfrac{1}{2}, \quad \mbox{if } j=k, \nonumber \\
0, \quad \mbox{if } j \ne k
\end{cases}
\end{align}
leads to the equality  
\begin{align}
\int_0^1 W^2(t) dt = \frac{1}{\pi^2} \sum_{k=1}^\infty \frac{Z_k^2}{(k-\tfrac{1}{2})^2}.
\end{align}
Further terms in the expansion can also be computed, albeit with more effort. For example, if we look at the cubic term, then 
\begin{equation}
\begin{split}
    &\int_0^1 \sin\Big((k_1-\tfrac{1}{2})\pi s \Big) \sin\Big( (k_2-\tfrac{1}{2})\pi s \Big) \sin\Big( (k_3-\tfrac{1}{2})\pi s \Big) ds \\
    &= \frac{1}{2\pi}\bigg(\frac{1}{3-2(k_1+k_2+k_3)}+\frac{1}{-1+2(k_1+k_2-k_3)} \\
    &\quad +\frac{1}{-1+2(k_1-k_2+k_3)}+\frac{1}{-1+2(-k_1+k_2+k_3)}\bigg) \\
    &=: \mathfrak{W}(k_1,k_2,k_3). \nonumber
\end{split}
\end{equation}
We note how the combinatorics becomes much complicated and all three indices $k_1, k_2$ and $k_3$ survive. Therefore, the next term in the expansion will given by
\begin{equation}
\begin{split}
    &\int_0^1 W^3(t)dt \nonumber \\
    &= \sumthree_{k_1, k_2, k_3 \ge 1} \frac{Z_{k_1}Z_{k_2}Z_{k_3}}{(k_1-\frac{1}{2})(k_2-\frac{1}{2})(k_3-\frac{1}{2})}\mathfrak{W}(k_1,k_2,k_3).
\end{split}
\end{equation}
This means that whereas in the quadratic case we were summing `shifted and re-scaled' squares of normal distributions, now we are summing even more shifted and re-scaled products of three normal distributions. This process can be carried out further, but two or three terms in the MacLaurin expansion intuitively suffices to obtain an adequate approximation. The remaining step is to add these normal distributions and obtain the emergent probability density function (PDF) of the new random variable. It is known, for example, that adding squares of normal distributions leads to the chi-squared($\chi^2$) distribution, i.e. if $Z_1, Z_2, \cdots, Z_k$ are i.i.d. standard normal variables, then
\begin{align}
    Q(k) = \sum_{n=1}^k Z_n^2 \sim \chi^2(k),
\end{align}
where $\chi^2$ has pdf given by
\begin{align}
    f_{\chi^2}(x) = \frac{x^{k/2-1}e^{-x/2}}{2^{k/2}\Gamma(k)}.
\end{align}
Unfortunately, this is not our situation since each $Z_k^2$ in our expansion for $\displaystyle \int_0^1 W^2(t)dt$ is affected by the coefficient $\displaystyle \frac{1}{(k-\frac{1}{2})^2}$. 

In addition, we are dealing with an infinite sum of i.i.d. normal random variables, not with a finite sum. Therefore, a truncation will be needed. In order to overcome the former situation, one needs to look at generating functions. Suppose we have two independent random variables $X$ and $Y$ with PDFs $f_X(x)$ and $f_Y(y)$ and set $Z=X+Y$. One can find the PDF $f_Z(x)$ of $Z$ by computing the convolution
\begin{align}
    f_Z(x) = \int_{-\infty}^\infty f_X(x)f_Y(z-x)dx.
\end{align}
In our case of interest, these convolutions would become increasingly difficult to solve analytically, requiring numerical methods to be utilized. Surprisingly, very little is known about how to add log-normal distributions. It is known that the \textit{product} of two log-normals is another log-normal distributions. However, addition has eluded an analytical solution. 

In summary, one can manufacture an approximation to the probability distribution of the Asian option by integrating the exponential of Brownian motion. This entails convolving powers of i.i.d. standard normal variables affected with certain algebraic coefficients and then adding the resulting PDFs. While this is a mathematically difficult problem, it bypasses the necessity of employing high dimensionality techniques as often invoked in the literature of quantum option pricing.

\section{Amplitude Estimation}

Of the steps required to price a derivative on a quantum computer, it is only in the last step of extracting the expected value where the quadratic speedup is obtained.  The retrieval of the expected value of these computations rely on Quantum Amplitude Estimation, which converges as $O(1/M)$ compared to the classical convergence rate of $O(1/\sqrt{M})$, as shown by \cite{montanaro}.  
Therefore there is a motivation to perform Amplitude Estimation (AE) in an efficient and accurate way, and to benchmark its success on current hardware.  Formally, the goal of AE is to provide an approximation of $a$ prepared by an operator $\mathcal{A}$, for instance, 
\begin{equation}
\mathcal{A}\ket{0_n}\ket{0} = \sqrt{1-a}\ket{\psi_0}_n\ket{0} + \sqrt{a}\ket{\psi_1}_n\ket{1}
\end{equation}

In other words, there is a set of quantum operators $\mathcal{A}$ (e.g., the payoff function of a call option) that maps a quantity of interest (e.g., the expected value of the option) into a quantum state $\ket{\psi_1}\ket{1}$ where the goal is to find the amplitude of that state, $\sqrt{a}$.
The canonical method was to use Quantum Phase Estimation (QPE) to provide an estimate for the amplitude \cite{tapp}.  

The QPE approach relied on using a controlled version of the Grover to iterate an oracle $\mathcal{Q}$ (Grover Iterate $\mathcal{Q}$), where $\mathcal{Q}$ is defined as 
\begin{equation*}
\mathcal{Q}= \mathcal{A}\mathcal{S}_0\mathcal{A^\dagger}\mathcal{S}_\psi.
\end{equation*}
Here $\mathcal{S}_\psi = \mathcal{I} - 2\ket{\psi_1}_n\bra{\psi_1}_n \bigotimes \ket{1}\bra{1}$ and $\mathcal{S}_0 = \mathcal{I} - 2\ket{0}_{n+1} \bra{0}_{n+1} $.  In other words, $\mathcal{S}_\psi$ adds a negative sign to the $\ket{\psi_1}$ states and $\mathcal{S}_0$ reflects around the zero state. Finally, performing Quantum Fourier Transform (QFT) on the control register of size $m$ we would get an estimate for $a$.  

 \begin{figure}[!t]
\centering
\begin{quantikz}[column sep=5pt, row sep={18pt,between origins}]
\lstick{ $\ket{0}^{n}$ } & \qw & \gate[2,nwires=2]{ \mathcal{A}  } & \gate[2,nwires = 2]{ \mathcal{Q}^{2^0} } & \gate[2,nwires = 2]{ \mathcal{Q}^{2^1}  } & \ \ldots\  \qw & \gate[2,nwires = 2]{ \mathcal{Q}^{2^{n-1} }  } & \qw & \qw
\\ \lstick{ \ket{0} \hspace{1.5pt} } & \qw & \qw & \qw  & \qw & \ \ldots\  \qw & \qw & \qw & \qw
\\ \lstick{(0) \ \ket{0}} & \gate{H} & \qw & \ctrl{-2}  & \qw & \ \ldots\  \qw & \qw & \gate[4,nwires = 3]{ \mathcal{F}_{n}^{\dagger}  } & \meter{}
\\ \lstick{(1) \ \ket{0}} & \gate{H} & \qw   & \qw & \ctrl{-2} & \ \ldots\   \qw  & \qw  & \qw & \meter{}
\\ & \vdots  &&&& \ddots &&& \vdots
\\ \lstick{(n-1) \ \ket{0}} & \gate{H} & \qw   & \qw  & \qw & \ \ldots\ \qw & \ctrl{-4} & \qw & \meter{}
\end{quantikz}
    \caption{Canonical algorithm for Amplitude Estimation by Quantum Phase Estimation.}
    \label{fig:qpe_ae}
\end{figure}
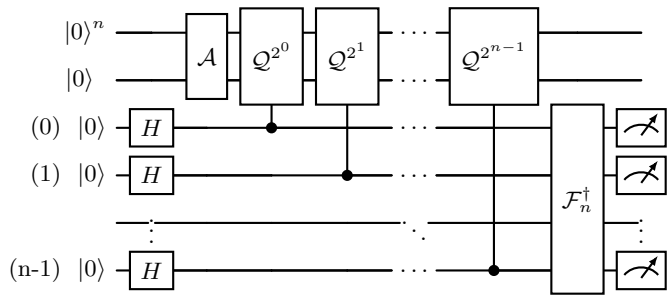

This method has operators that make it unsuitable for current noisy devices. The qubit topography of many quantum systems make control gates expensive and can quickly increase the depth of the circuit.  Similarly, QFT requires expensive operations that are generally not practically implementable on current devices.

There has been active research into other ways to conduct AE, some even involving variational methods \cite{plekhanov2021variational}.  Most current methods involve non-controlled applications of $\mathcal{Q}$ and some classical post processing to create an estimate of $a$ \cite{grinko2019iterative,S20}.  These methods rely on the fact that $\mathcal{A}\ket{0_n} = \cos{\theta}\ket{\psi_0} + \sin{\theta} \ket{\psi_1}$.  If we can get an accurate estimate for $\theta$, then $\sin^2{\theta} = a$.  Included in these methods are the Maximum Likelihood Amplitude Estimation (MLAE) algorithm \cite{S20}, which applies successive applications of the Grover Iterate $\mathcal{Q}$ and finds the $\theta$ that is most likely given the results of the circuit.  A benefit of MLAE is that it allows us to control the number of applications, as well as the number of circuit evaluations for the $\mathcal{Q}$ oracle.

Given a number $m_k$ of repetitions of $\mathcal{Q}$, the number of circuit evaluations $N_k$, and the number of times $h_k$ that $\ket{\psi_1}\ket{1}$ was measured, the probability of measuring $\ket{\psi_1}\ket{1}$ is $P(m_k;a) = \sin^2\Big( (2m_k +1)\theta_a \Big)$.  The likelihood function is therefore

\begin{equation}
\mathcal{L}(h;a)= \prod_{k=0}^M P(m_k;a)^{h_k}[1-P(m_k;a)]^{N_k-h_k}
\end{equation}

The number of repetitions of $\mathcal{Q}$, as well as the number of circuit evaluations per each repetition define the sequencing schedule for the algorithm.  Different schedules have been proposed, including the Linearly Incremental Sequence (LIS), the Exponential Incremental Sequence (EIS), and the Power Law schedule \cite{tanaka2021amplitude, giurgica2022low}.

\section{Linear Payoff Function}

\begin{figure*}[t]
    \centering
    \includegraphics[width=200px] {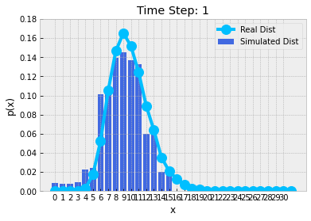}  
    \includegraphics[width=200px]{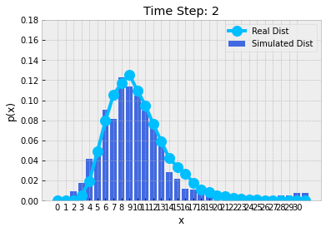}
    \includegraphics[width=200px]{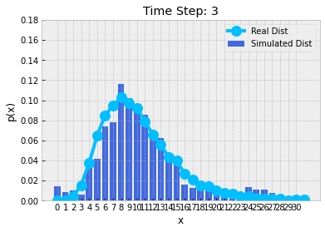}
    \includegraphics[width=200px]{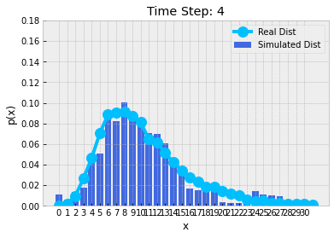}
    \caption{The histograms above show the results of the qGAN on each of the time steps.}
    \label{fig:results_combined}
\end{figure*}

The payoff function of the Asian option contract we have introduced is piece-wise linear, and thus can be constructed from the method in \cite{S19}.  We first incorporate an integer comparison operator, which flips a target qubit if the basis state is greater than or equal to an integer:

\begin{equation*}
\ket{i}_{n} \ket{0} \mapsto \ket{i}_{n} \ket{i \geq L}
\end{equation*}

Here we choose L to be the mapped basis state of the strike price.  Using this ancilla qubit and the distribution register, we can now use a series of controlled-y rotations to embed the piecewise linear function into a final qubit, where AE is to be performed.  Due to the nonlinear change of the amplitude of the qubit to multiple rotational gates, we center the amplitude around $\sin{\frac{\pi}{4}}$ and use a small value $c$ so that interval of the range of the function is $2c$. Hence, the minimum of the function is mapped to the amplitudes to $\sin{(\frac{\pi}{4} - c)}$ and the maximum to $\sin{(\frac{\pi}{4} + c)}$.

\begin{figure}[H]
    \centering
    \includegraphics[width=240pt]{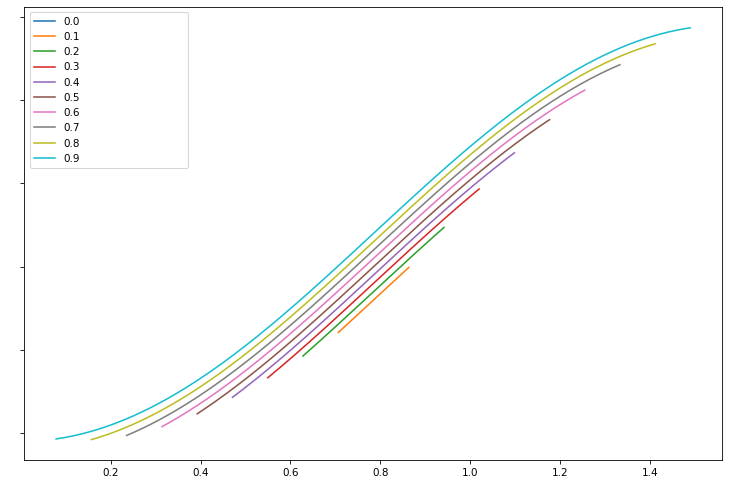}
    \caption{The figure displays how well different values of $c$ approximate `close-to' linear part of the sine function. Observer that smaller values of $c$ are better estimates.}
    \label{fig:sine-c}
\end{figure}

The smaller $c$ values lead to centering the payoff function around the `close-to linear' part of sine, and will give a more accurate value when extracted with AE, albeit with slightly worse convergence.  This can be shown by plotting the different intervals of the AE of the qubit amplitudes given a value of c, shown in Figure \ref{fig:sine-c}.

In \cite{herbertquantum}, Herbert demonstrated that the economy from the arithmetic of the payoff function cannot ever be quadratic within the accepted mainframe presented in \cite{egger2019credit, GS}. Therefore, the technique present in \cite{GS} has an adverse effect on the rate of convergence, as the requirement of quantum arithmetic was removed to the reduce the circuit depth, but comes at the cost of reduced convergence rate. This implies that deeper circuits will eventually be needed to achieve the desired estimation accuracy. Herbert solves this problem by employing a Fourier series decomposition of the sum that approximates the payoff of the European option, where each component is estimated individually using QAE. Adapting Hebert's work to the payoff of the Asian option will be the objective of future research.

\section{Results}\label{sec:results}

To demonstrate the effectiveness of our approach, we first classically simulated the evolution of a stock price over 4 time steps with a risk free rate of $.1$ and annualized volatility of $.5$ and an initial price of $100$.  The resulting distributions had a minimum value of $17.$ and a maximum value of $300.$, of which we discretized with 32 intervals (5 qubits).  After training, the distributions for each time step can be seen in Figure \ref{fig:results_combined}.  

\begin{figure}[!t]
    \centering
    \includegraphics[width=255pt]{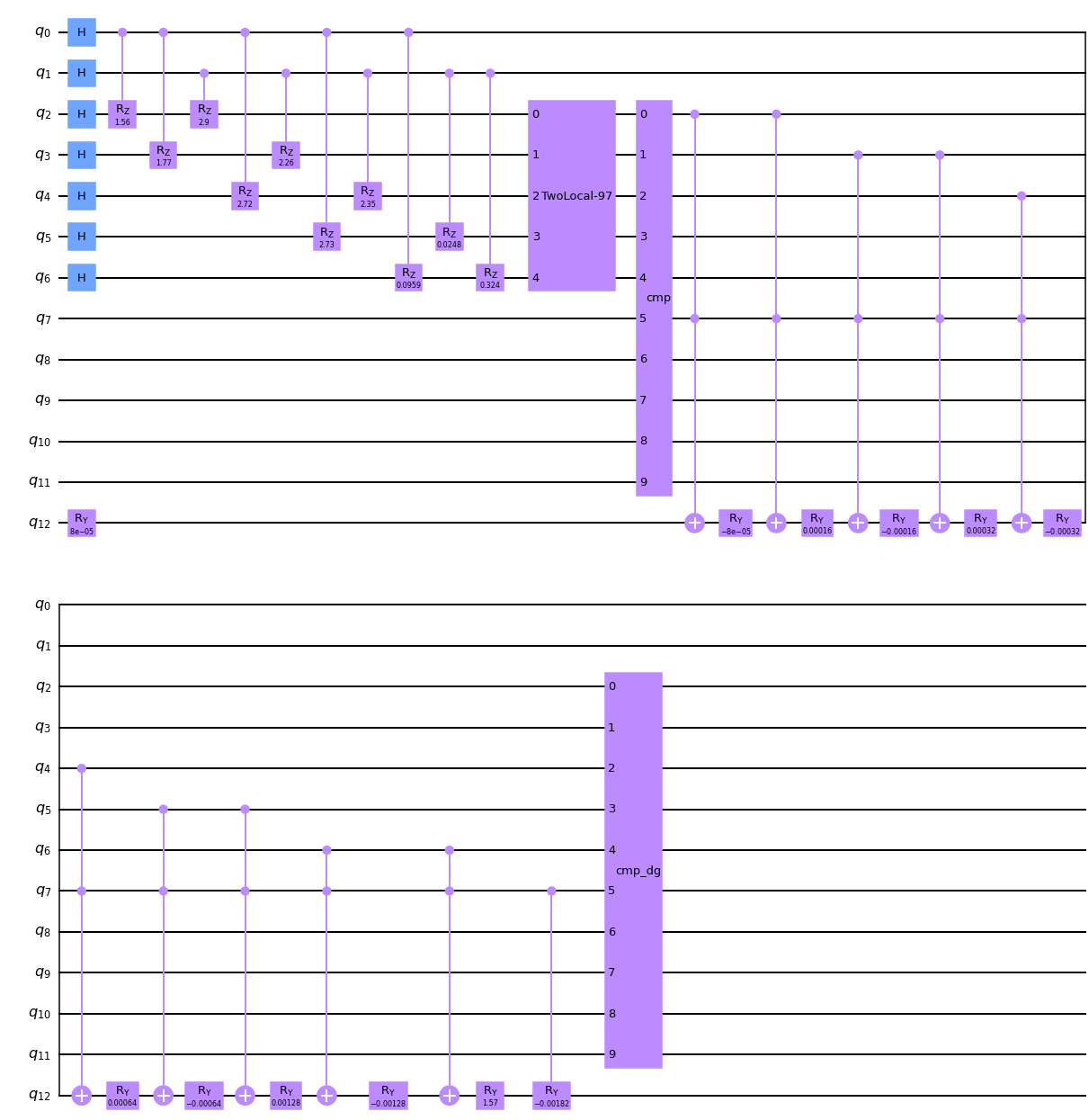}
    \caption{Option pricing circuit incorporating C-qGAN as a subprocess.}
    \label{fig:ae_opt_cir}
\end{figure}

Therefore, the steps to price the Asian option is to first initialize these time steps in equal superposition, creating a simple average of them in the distribution register, then performing the integer comparator, followed by our linear payoff functions via controlled-y rotations, before finally extracting the expected value of the last qubit through AE.  The full circuit is shown in Figure \ref{fig:ae_opt_cir}.

With a strike price of $110.$, the true value of the Asian option via the discretized distributions is $13.854$ vs. $14.41$, the value found from the distributions actually loaded on the quantum simulator.  This difference represents the error in our training regime, of which we believe there is room for improvement to narrow this gap.  Our quantum circuit in Figure \ref{fig:ae_opt_cir}, given an appropriately small $c$ value, arrives at the correct fair value price of $14.41$.  

\section{Code availability}
The code that supports the findings of this study is available from the corresponding author SC upon reasonable request.

\section{Acknowledgment}
The authors would like to acknowledge participation in the IBM Quantum Finance Working Group Kickoff that took place in October 2022 in New York City where the ideas for this manuscript were first discussed.

\section{Disclaimer}
\subsection{Deloitte}
About Deloitte: Deloitte refers to one or more of Deloitte Touche Tohmatsu Limited (“DTTL”), its global network of member firms, and their related entities (collectively, the “Deloitte organization”). DTTL (also referred to as “Deloitte Global”) and each of its member firms and related entities are legally separate and independent entities, which cannot obligate or bind each other in respect of third parties. DTTL and each DTTL member firm and related entity is liable only for its own acts and omissions, and not those of each other. DTTL does not provide services to clients. Please see www.deloitte.com/about to learn more.

Deloitte is a leading global provider of audit and assurance, consulting, financial advisory, risk advisory, tax and related services. Our global network of member firms and related entities in more than 150 countries and territories (collectively, the “Deloitte organization”) serves four out of five Fortune Global 500® companies. Learn how Deloitte’s
approximately 330,000 people make an impact that matters at www.deloitte.com. 
This communication contains general information only, and none of Deloitte Touche Tohmatsu Limited (“DTTL”), its global network of member firms or their related entities (collectively, the “Deloitte organization”) is, by means of this communication, rendering professional advice or services. Before making any decision or taking any action that
may affect your finances or your business, you should consult a qualified professional adviser. No representations, warranties or undertakings (express or implied) are given as to the accuracy or completeness of the information in this communication, and none of DTTL, its member firms, related entities, employees or agents shall be liable or
responsible for any loss or damage whatsoever arising directly or indirectly in connection with any person relying on this communication. 
Copyright © 2022. For information contact Deloitte Global.

\subsection{IBM}
IBM, the IBM logo, and ibm.com are trademarks of International Business Machines
Corp., registered in many jurisdictions worldwide. Other product and service names might be trademarks of IBM or other companies. The current list of IBM trademarks is available at https://www.ibm.com/legal/copytrade.

\bibliographystyle{unsrt}
\bibliography{refs}

\newpage
\appendix
\renewcommand\thefigure{\thesection.\arabic{figure}}
\setcounter{figure}{0}    
~\\
\section{Training Conditional-qGAN to Load Multiple Distributions on a Noisy Device}

\begin{figure*}[!b]
    \centering
    \includegraphics[width=400pt]{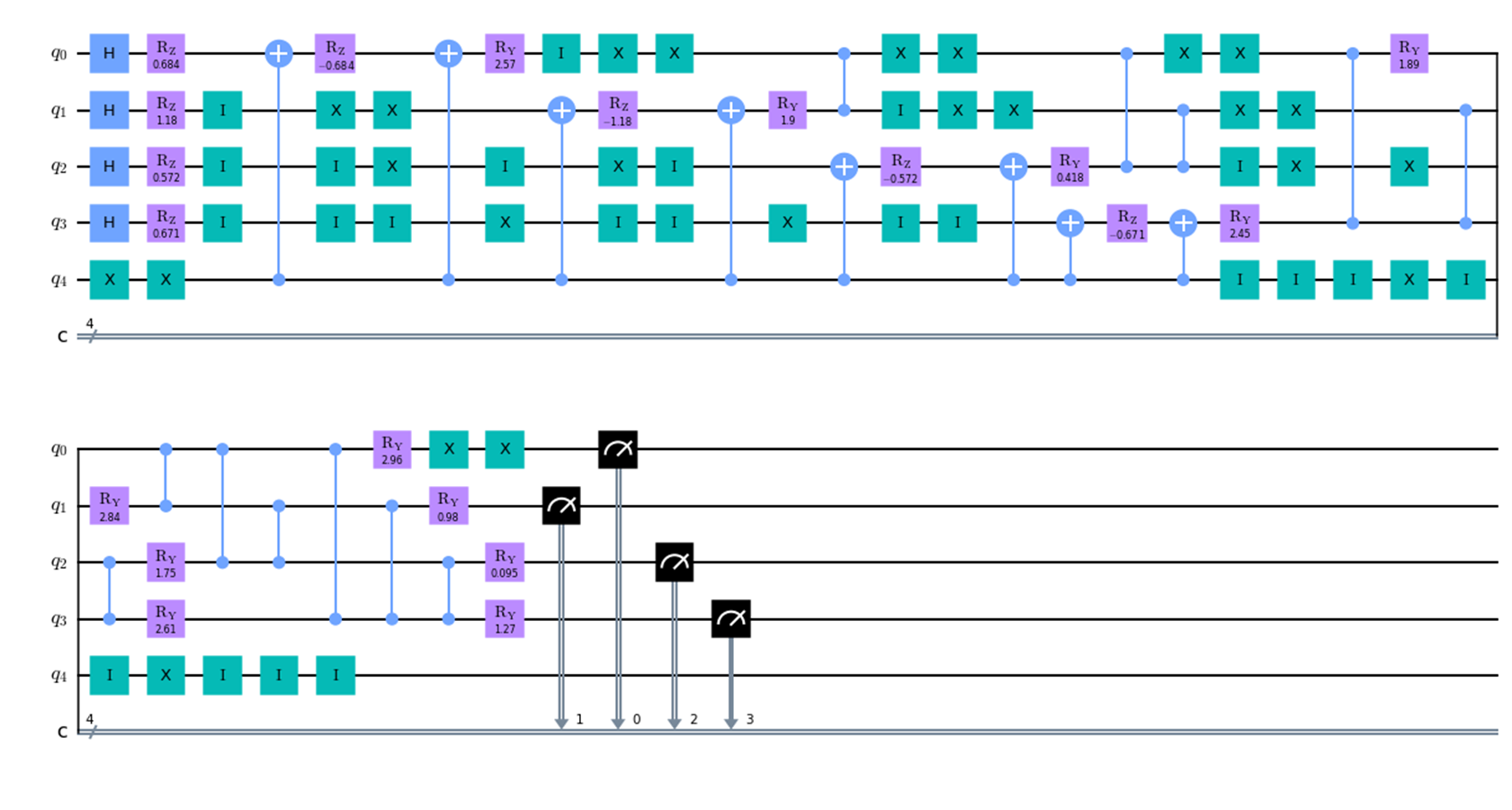}
    \caption{C-qGAN circuit with dynamic decoupling using $X$ gates.  Here $q_{4}$ is the conditional qubits, and $q_{0}$ through $q_{3}$ outputs the discretized distribution with $2^{4}$ states.}
    \label{fig:cqan_dd}
\end{figure*}

\begin{figure*}[!b]
    \centering
    \includegraphics[width=350pt]{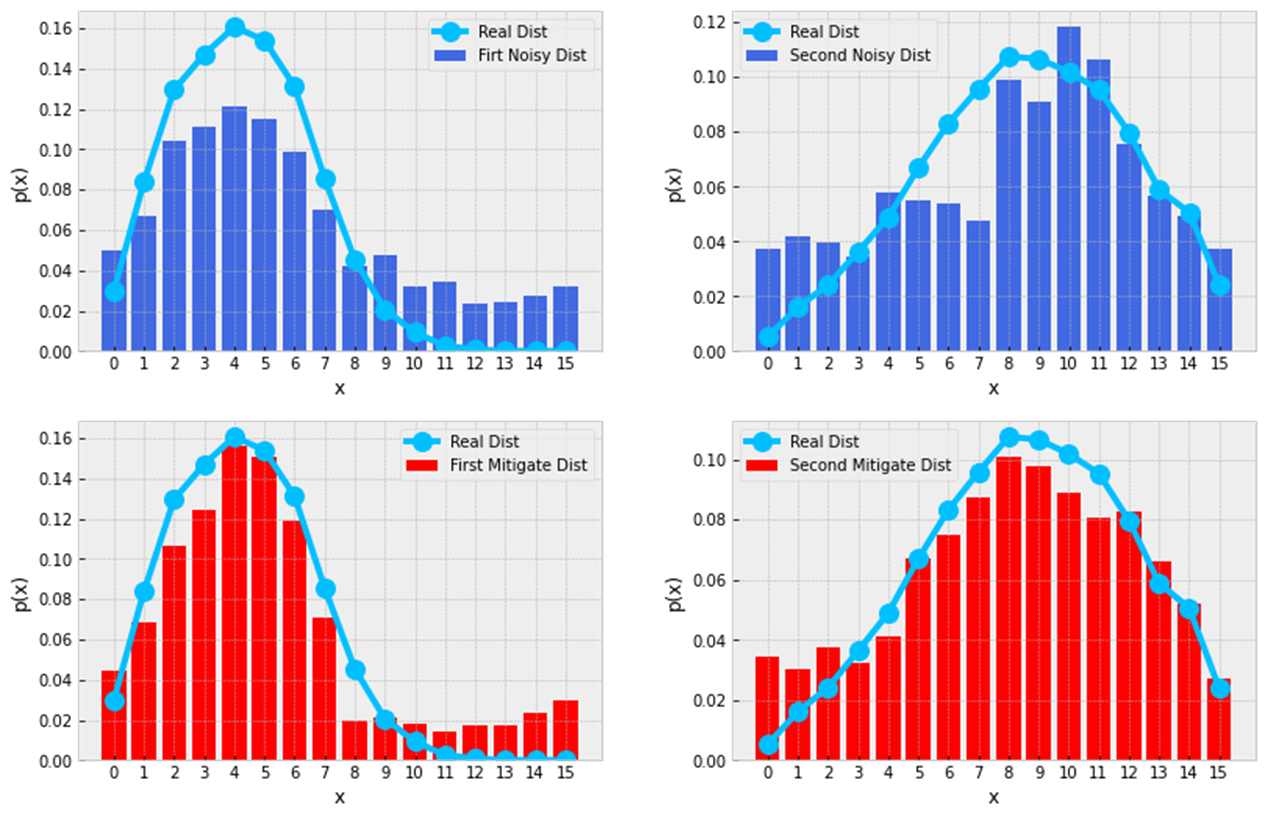}
    \caption{Distributions generated without error mitigation (blue) and with error mitigation (red) using our C-qGAN model.}
    \label{fig:cqan_dist_dd}
\end{figure*}

A previous study suggested the feasibility of loading a random probability distribution using a noisy device \cite{Zoufal_2019} due to the short circuit depth of the ansatz in the qGAN model. To explore the possibility of loading multiple distributions using our C-qGAN on a noisy device, we conducted a simple experiment to train our C-qGAN circuit with 4 qubits and one conditional qubit to load two random distributions. For our experiment, we used the simulated Nairobi backend which contains the same topology and noise model as the 7-qubit IBMQ Nairobi device. Specifically, we used 10000 shots and utilized the gradient free SPSA classical optimizer \cite{spall2005introduction}. Error mitigation using dynamic decoupling technique was also implemented using the Mitiq package \cite{LaRose2020rz}. Dynamic decoupling adds a series of simple pauli gates to the idle qubits to minimize the effect of crosstalk errors \cite{Viola2417}. A simple application of the pauli $X$ gates [Fig. \ref{fig:cqan_dd}] to the idle qubits in our C-qGAN model produces improved training results to load the two targeted probabilities [Fig. \ref{fig:cqan_dist_dd}].

\onecolumngrid

\end{document}